\documentclass[sigconf, nonacm]{acmart}

\newcommand\vldbavailabilityurl{https://github.com/lixiang-lab/benchmark_grnnd}

\usepackage{tabularx}
\usepackage{booktabs}
\usepackage{subcaption}
\usepackage{algorithm}
\usepackage{algorithmic}
\usepackage{amsmath}
\usepackage{diagbox}
\usepackage{multirow}
\usepackage{xcolor}
\usepackage{tikz} 
\usepackage{pgfplots}
\pgfplotsset{compat=1.18}
\usepgfplotslibrary{groupplots}
\usepackage{pgfplotstable}
\usepackage{threeparttable}
\usepackage{setspace}
\usepackage{enumitem}
\usetikzlibrary{decorations.pathmorphing}
\pgfplotsset{
  compat=1.18,
  jumpy/.style={
    unbounded coords=jump,
    filter discard warning=false
  },
  lowseg/.style={restrict y to domain*=-1:40},
  highseg/.style={restrict y to domain*=70:100000, forget plot},
}

\begin{document}
\title{GRNND: A GPU-Parallel Relative NN-Descent Algorithm for Efficient Approximate Nearest Neighbor Graph Construction}

\author{Xiang Li}
\orcid{0000-0001-6933-6491}
\affiliation{
  \institution{Nanjing University}
  \city{Nanjing}
  \country{China}
}
\email{652022230010@smail.nju.edu.cn}

\author{Qiong Chang}
\authornote{Corresponding author.}
\orcid{0000-0002-4447-0480}
\affiliation{
  \institution{Institute of Science Tokyo}
  \city{Tokyo}
  \country{Japan}
}
\email{q.chang@c.titech.ac.jp}

\author{Yun Li}
\orcid{0000-0003-1753-7317}
\affiliation{
  \institution{Nanjing University}
  \city{Nanjing}
  \country{China}
}
\email{yli@nju.edu.cn}

\author{Jun Miyazaki}
\orcid{0000-0002-3038-7678}
\affiliation{
  \institution{Institute of Science Tokyo}
  \city{Tokyo}
  \country{Japan}
}
\email{miyazaki@c.titech.ac.jp}

\begin{abstract}
Relative Nearest Neighbor Descent (RNN-Descent) is a state-of-the-art algorithm for constructing sparse approximate nearest neighbor (ANN) graphs by combining the iterative refinement of NN-Descent with the edge-pruning rules of the Relative Neighborhood Graph (RNG). It has demonstrated strong effectiveness in large-scale search tasks such as information retrieval and related tasks. 
However, as the amount and dimensionality of data increase, the complexity of graph construction in RNN-Descent rises sharply, making this stage increasingly time-consuming and even prohibitive for subsequent query processing.
In this paper, we propose GRNND, the first GPU-parallel algorithm of RNN-Descent designed to fully exploit GPU architecture. GRNND introduces a disordered neighbor propagation strategy to mitigate synchronized update traps, enhancing structural diversity, and avoiding premature convergence during parallel execution. It also leverages warp-level cooperative operations and a double-buffered neighbor pool with fixed capacity for efficient memory access, eliminate contention, and enable highly parallelized neighbor updates. Extensive experiments demonstrate that GRNND consistently outperforms existing CPU- and GPU-based methods. GRNND achieves $2.4$ to $51.7\times$ speedup over existing GPU methods, and $17.8$ to $49.8\times$ speedup over CPU methods.
\end{abstract}


\maketitle

\begingroup\small\noindent\raggedright\textbf{Reference Format:}\\
Author(s). Title. Preprint, arXiv:\href{https://arxiv.org/abs/xxxx.xxxxx}{xxxx.xxxxx}, Year.
\endgroup

\begingroup
\renewcommand\thefootnote{}\footnote{\noindent
This is a preprint of an article submitted to arXiv. 
The copyright of this work remains with the author(s). 
Licensed under a Creative Commons BY-NC-ND 4.0 International License. 
Visit \url{https://creativecommons.org/licenses/by-nc-nd/4.0/} to view a copy of this license. 
}
\addtocounter{footnote}{-1}
\endgroup

\ifdefempty{\vldbavailabilityurl}{}{
\vspace{.3cm}
\begingroup\small\noindent\raggedright\textbf{Artifact Availability:}\\
The source code, data, and/or other artifacts are available at \url{\vldbavailabilityurl}.
\endgroup
}

\section{Introduction}
Approximate Nearest Neighbor Search (ANNS) is a fundamental technique for large-scale high-dimensional data analysis, with widespread applications in information retrieval\cite{li2019approximate,huang2020embedding,vecchiato2024learning,li2020improving,gupta2022bliss}, natural language processing\cite{wang2021comprehensive,xiong2020approximate,tu2020approximate}, and computer vision\cite{carrara2022approximate,harwood2016fanng,baranchuk2018revisiting}. By tolerating a small accuracy loss, ANNS provides significant gains in scalability and search efficiency.
In particular, graph-based ANNS methods stand out for their excellent trade-off between accuracy and speed, relying on a pre-constructed proximity graph for efficient querying.
NN-Descent~\cite{dong2011efficient} represents a foundational method for constructing approximate $k$-NN graphs via iterative refinement. 
Based on it, two main strategies have emerged. The first builds a dense $k$-NN graph and applies pruning to obtain a sparser structure, such as NSG~\cite{fu2017fast}, SSG~\cite{fu2021high}, and CAGRA~\cite{ootomo2024cagra}. The second integrates pruning directly into the construction process to obtain sparse graphs from the outset, as in NSW~\cite{malkov2014approximate} and its GPU variant GANNS~\cite{yu2022gpu}, HNSW~\cite{malkov2018efficient} and RNN-Descent~\cite{ono2023relative}, thereby reducing the additional computation overhead introduced by dense pre-construction and subsequent pruning.

Within the direct-construction family, NSW offers simplicity and good navigability, but relies on randomized insertions that can produce uneven degrees and quality variance at scale. HNSW improves accuracy and robustness through hierarchical layering and heuristic pruning, but its incremental, order-dependent insertions and dynamic memory use introduce strong sequential dependencies and irregular memory behavior, which hinder parallelization. RNN-Descent, recently proposed by Ono et al.~\cite{ono2023relative}, is the current state-of-the-art for direct graph construction. Integrates the iterative propagation strategy of NN-Descent with edge pruning based on the Relative Neighborhood Graph (RNG) criterion~\cite{jaromczyk2002relative}. This combination enables the construction of high-quality sparse graphs and consistently outperforms previous methods across multiple benchmarks. 

%

With increasing amount and dimensionality of data, the cost of ANN graph construction grows rapidly, largely attributed to high-dimensional distance evaluations and neighbor maintenance tasks such as candidate sorting, deduplication, and redirection. Consequently, construction time often becomes the dominant stage in the end-to-end ANN retrieval pipeline, delaying or even preventing subsequent query processing~\cite{ruan2025empowering,liu2022optimizing,luo2025efficient}.
In this context, Graphics Processing Units (GPUs) offer massive parallelism and memory bandwidth~\cite{jin2020understanding,liu2025smore,gale2020sparse}, and have shown great promise in accelerating the online search stage of ANNS~\cite{wieschollek2016efficient,zhao2020song,Karthik2025BANG}. However, translating construction to GPUs has yielded limited gains without carefully structured algorithms~\cite{chang2025accelerating,li2024optimized,chang2023multi,Li20253DGNLM}, and as a result, efficient graph construction on GPUs remains challenging~\cite{wang2025accelerating,meyer2021warp}. 
In particular, the original design of RNN-Descent assumes sequential iteration, ascending-order neighbor updates, and fine-grained dynamic memory access, which conflict with the GPU execution model. Direct parallelization leads to premature convergence: vertices update neighbors in the same ascending order without feedback, causing unnecessary reinsertion, dense local connections, and suppressed exploration. These issues prevent effective pruning and trap the graph in low-quality states. 

To overcome these limitations, we propose GRNND, the first GPU-parallel algorithm of RNN-Descent. GRNND retains the core iterative refinement principles of RNN-Descent, but systematically redesigns its execution strategy and memory organization to align with the GPU architecture. It targets three key challenges: (1) avoiding premature convergence, (2) replacing dynamic memory with a fixed-capacity double-buffered neighbor pool, and (3) reducing control divergence and memory contention. 
The contributions of this paper can be summarized as follows:
\begin{itemize}
\item \textbf{Disordered neighbor propagation}: Instead of updating in ascending order, GRNND evaluates the candidate neighbors in randomized order, which avoids premature convergence, reduces unnecessary reinsertion, and promotes structural diversity. 
\item \textbf{Warp-level cooperative operation}: Each vertex is updated by a single warp using warp-level primitives (e.g., ballot, shuffle), enabling efficient distance computation and insertion with minimal control divergence and no global synchronization.
\item \textbf{Fixed-capacity double-buffered pool}: Dynamic allocations are replaced by two fixed-size buffers per vertex, ensuring lock-free updates, efficient memory access, and natural alignment with warp-level operations.
\item \textbf{Reverse edge sampling}: Rather than inserting reverse edges exhaustively, GRNND samples a subset each iteration, reducing computation while maintaining graph quality and exploration.
\end{itemize}
On RTX 4090, GRNND delivers $2.6$ to $48.3\times$ speedup over GPU baselines and $27.1$ to $49.8\times$ over CPU baselines.
On RTX 6000 Ada, the corresponding ranges are $2.4$ to $51.7\times$ (GPU) and $17.8$ to $26.2\times$ (CPU), confirming consistent advantages across hardware platforms.

The remainder of this paper is organized as follows. Section~\ref{sec_background} presents background information. Section~\ref{sec_implementation} describes our GPU-parallel algorithm. Section~\ref{sec_experiment} reports the experimental evaluation. Section~\ref{sec_releatedworks} reviews related work and Section~\ref{sec_conclusion} concludes the paper.
\begin{table}[ht]
\centering
\caption{Notation used throughout the paper.}
\label{tab:notation}
\begin{tabularx}{\columnwidth}{lX}
\toprule
\textbf{Symbol} & \textbf{Meaning} \\
\midrule
$V$ & Vertex set (dataset) \\
$E$ & Edge set \\
$v \in V$ & A vertex \\
$n$ & A neighbor of $v$ \\
$N_v$ & Candidate neighbor set of $v$ \\
$N_v'$ & Refined neighbor set of $v$ \\
$d(v_1, v_2)$ & Distance function between vectors $v_1$ and $v_2$ \\
$S$ & Number of initial neighbors per vertex \\
$R$ & Pool capacity (maximum neighbors per vertex) \\
$T_1$ & Number of outer iterations \\
$T_2$ & Number of inner iterations \\
$\rho$ & Reverse-edge sampling ratio \\
\bottomrule
\end{tabularx}
\end{table}

\section{BACKGROUND}
\label{sec_background}
For clarity, the notations used throughout this paper are summarized in Table~\ref{tab:notation}. 
\subsection{Approximate Nearest Neighbor Search}
The Approximate Nearest Neighbor Search (ANNS) aims to retrieve vectors close to a query $\mathbf{q} \in \mathbb{R}^d$ by trading exactness for improved efficiency. A typical ANNS system consists of two stages: index construction and online querying.
Graph-based methods have emerged as one of the most effective approaches due to their superior trade-off between accuracy and efficiency~\cite{wang2021comprehensive}. They are based on a navigable proximity graph, where the quality of the graph largely determines search performance. Sparse graphs with adaptive neighbor sets often support more efficient navigation than uniform $k$-NN graphs, but their construction is computationally costly.
During querying, performance is usually evaluated using two metrics: Recall@k and Queries Per Second (QPS). Recall measures accuracy as the fraction of true nearest neighbors retrieved, while QPS quantifies the number of queries processed per unit of time, indicating the efficiency of the search process.

\subsection{RNN-Descent Algorithm}
\begin{algorithm}
\caption{RNN-Descent Algorithm}
\label{alg:rnn_descent}
\begin{algorithmic}[1]
\STATE \textbf{Input:} $V$, $N_v$, $S$, $R$, $T_1$, $T_2$
\STATE \textbf{Output:} Graph$(V,E)$
\STATE $\forall v \in V$: $\text{INITIALIZATION}(v, S, N_v)$
\FOR{$t_1 = 1$ to $T_1$}
    \FOR{$t_2 = 1$ to $T_2$}
        \STATE $\forall v \in V$: $\text{UPDATE\_NEIGHBORS}(v, R, N_v)$ 
    \ENDFOR
    \IF{$t_1 \neq T_1$}
        \STATE $\forall v \in V$: $\text{ADD\_REVERSE\_EDGES}(v, R, N_v)$ 
    \ENDIF
\ENDFOR
\end{algorithmic}
\end{algorithm}

RNN-Descent begins with a randomly initialized $S$-NN index and iteratively refines the graph by applying the RNG criterion. For a vertex $v$ with two candidate neighbors $n_1$, $n_2$, both neighbors are retained only if:
\begin{equation}
    \begin{aligned}
        d(n_1, v) < d(n_1, n_2) \land d(n_2, v) < d(n_1, n_2).
    \end{aligned}
\label{eq:rng}
\end{equation}

\begin{figure}[ht]
  \centering
  \includegraphics[width=0.95\linewidth]{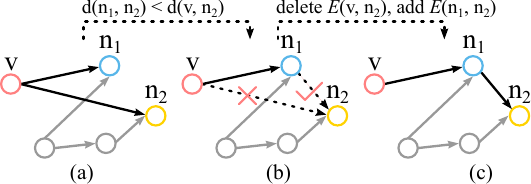}
  \caption{Neighbor refinement of vertex $v$ under the RNG criterion. (a) Initial candidate neighbors; (b) RNG-based edge pruning; (c) Refined neighbor set of $v$.}
  \label{fig:rng}
    \Description{Three subfigures illustrating the refinement of a neighbor graph using the Relative Neighborhood Graph (RNG) strategy. Subfigure (a) shows the initial graph before the neighbor update. Subfigure (b) highlights the edge selection process based on the RNG criterion, where an edge is retained only if no closer common neighbor exists. Subfigure (c) presents the updated graph after applying the RNG strategy, with spurious edges removed to improve graph quality.}
\end{figure}

In other words, if $n_1$ and $n_2$ are closer to each other than either is to $v$, then the edge between them takes precedence over their connections to $v$. As shown in Figure~\ref{fig:rng}, the vertex $v$ is initially connected to two neighbors $n_1$ and $n_2$. Since $n_2$ is closer to $n_1$ than to $v$, the edge between $v$ and $n_2$ is removed and replaced by an edge between $n_1$ and $n_2$.
The complete algorithm is presented in Algorithm~\ref{alg:rnn_descent} and consists of three key steps: (1) random initialization of neighbor pools (line 3), (2) iterative refinement based on the RNG criterion (line 6), and (3) insertion of reverse edges to maintain structural diversity and avoid premature convergence (line 9).

Each vertex sequentially refines its candidate list $N_v$, as detailed in Algorithm~\ref{alg:update_neighbors}. First, $N_v$ is sorted by ascending distance $d(v,n)$ and duplicates are removed (line 3). The first neighbor (i.e., closest) is always accepted and added to $N_v'$. 
Each subsequent candidate $n \in N_v$ (line 5) is compared with all previously accepted neighbors $n' \in N_v'$ (line 7). 
If any $n$ satisfies $d(n, n') \leq d(v, n)$ (line 8), $n$ is excluded from $N_v'$ and redirected to $N_{n'}$ (lines 9–11). Otherwise, $n$ is retained and added to $N_v'$ (line 15).

\begin{algorithm}
\caption{UPDATE\_NEIGHBORS}
\label{alg:update_neighbors}
\begin{algorithmic}[1]
\STATE \textbf{Input:} $v \in V$, candidate neighbors $N_v$
\STATE \textbf{Output:} Refined neighbors $N_v'$
\STATE Sort $N_v$ by ascending $d(v, n)$ and remove duplicates
\STATE Retain top $R$ closest candidates in $N_v$
\FOR{$n \in N_v$}
    \STATE $valid \leftarrow \texttt{true}$
    \FOR{$n' \in N_v'$}
        \IF{$d(n, n') \leq d(v, n)$}
            \STATE $valid \leftarrow \texttt{false}$
            \STATE $N_{n'} \leftarrow N_{n'} \cup \{n\}$
            \STATE \textbf{break}
        \ENDIF
    \ENDFOR
    \IF{$valid$}
        \STATE $N_v' \leftarrow N_v' \cup \{n\}$
    \ENDIF
\ENDFOR
\end{algorithmic}
\end{algorithm}

Over iterations, the graph may prematurely converge to suboptimal configurations due to the strict RNG constraint. To prevent this, RNN-Descent inserts reverse edges at the end of each outer iteration (except the last). These additional connections, which may violate the RNG criterion, introduce structural diversity and promote continued refinement in the next iteration.


\section{Parallel RNN-Descent on GPU}
\label{sec_implementation}
\begin{figure*}[ht]
    \centering
    \begin{subfigure}{0.9\linewidth}
        \centering
        \includegraphics[width=0.95\linewidth]{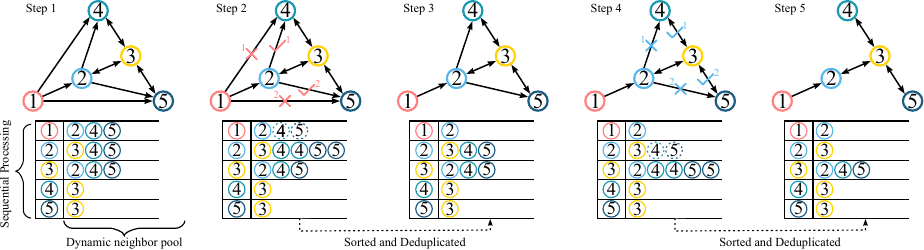}
        \caption{Serial iteration: neighbor updates occur sequentially, enabling stepwise propagation of structural information.}
        \label{fig:serial}
    \end{subfigure}
    \vspace{0.6em}

    \begin{subfigure}{0.9\linewidth}
        \centering
        \includegraphics[width=0.95\linewidth]{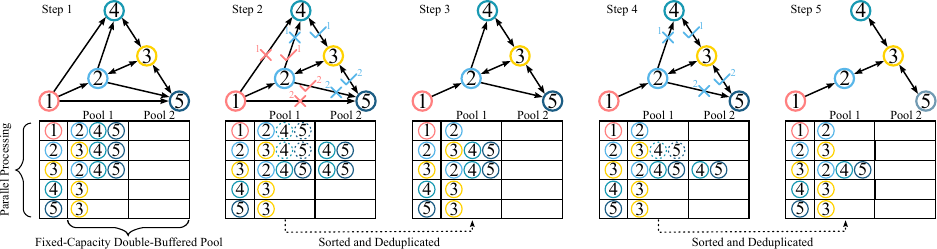}
        \caption{Parallel sorted iteration: each vertex independently updates in ascending order, breaking propagation chains and leading to local convergence traps.}
        \label{fig:parallel_sorted}
    \end{subfigure}
    \vspace{0.6em}

    \begin{subfigure}{0.9\linewidth}
        \centering
        \includegraphics[width=0.95\linewidth]{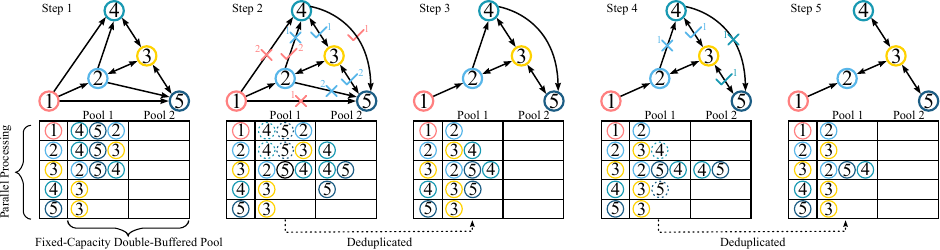}
        \caption{Parallel disordered iteration: candidate updates follow no fixed order, restoring stochastic propagation and enabling exploration beyond local optima.}
        \label{fig:parallel_disordered}
    \end{subfigure}

    \caption{
        Comparison of serial vs. parallel execution in RNN-Descent. While serial scheduling enables gradual global optimization, direct parallelism with sorted updates breaks propagation order and leads to premature local convergence. Disordered parallel updates restore exploration capability by injecting randomness into the update paths.
    }
    \label{fig:serial_parallel_compare}
    \Description{Three subfigures comparing different scheduling strategies in RNN-Descent. Subfigure (a) shows serial iteration, where vertices update one by one, allowing structural information to propagate step by step. Subfigure (b) shows parallel sorted iteration, where vertices update independently in ascending order, breaking propagation chains and causing local convergence. Subfigure (c) shows parallel disordered iteration, where vertex updates occur in a random order, enabling broader exploration and avoiding local traps.}
\end{figure*}

This section first analyzes the challenges of parallelizing RNN-Descent on GPUs and then presents our proposed GRNND.

\subsection{Challenges in Parallelizing RNN-Descent}
Although RNN-Descent is highly effective on CPUs, its neighbor update mechanism is inherently sequential, and thus unsuitable for direct GPU parallelization. Figure~\ref{fig:serial} illustrates this sequential process: Each vertex sorts and deduplicates its candidate neighbors, then iteratively applies the RNG criterion in ascending order.  
Take vertex 1 as an example. Its closest neighbor, vertex 2, is always retained. The remaining candidates, such as vertices 4 and 5, are evaluated against this retained set. For example, vertex 4 is closer to vertex 2 than to vertex 1, so it is excluded from the pool of vertex 1 and redirected to vertex 2.
This step-by-step process enables progressive refinement. Each vertex updates after others and can incorporate their structural improvements. Crucially, each candidate is compared to the closest neighbor retained in the central vertex. This relative evaluation ensures that redirection decisions are optimal locally and topologically coherent.
\begin{algorithm}
\caption{GRNND Algorithm}
\label{alg:grnnd}
\begin{algorithmic}[1]
\STATE \textbf{Input:} $V$, $N_v$, $S$, $R$, $T_1$, $T_2$, $\rho$
\STATE \textbf{Output:} $Graph(V,E)$
\FORALL{$v \in V$ \textbf{in parallel}}
    \STATE $\text{INITIAL\_NEIGHBORS}(v, S, N_v)$ 
\ENDFOR
\FOR{$t_1 = 1, 2, \ldots, T_1$}
    \FOR{$t_2 = 1, 2, \ldots, T_2$}
        \FORALL{$v \in V$ \textbf{in parallel}}
            \STATE \text{UPDATE\_NEIGHBORS\_PARALLEL}$(v, N_v)$ 
        \ENDFOR
    \ENDFOR
    \IF{$t_1 \neq T_1$}
        \FORALL{$v \in V$ \textbf{in parallel}}
            \STATE \text{REVERSE\_EDGE\_SAMPLING}$(v, N_v, \rho)$
        \ENDFOR
    \ENDIF
\ENDFOR
\end{algorithmic}
\end{algorithm}

This relative comparison, enabled by sequential updates, is central to RNN-Descent’s ability to refine the graph progressively on CPUs. However, when all vertices update their neighbors in parallel on GPUs, this sequential refinement is lost. In direct parallel implementation (Figure~\ref{fig:parallel_sorted}), vertices follow the same ascending-order rule and update independently without awareness of structural changes elsewhere in the graph. Such uniform update behavior leads to premature convergence, a failure mode characterized by three compounding effects.
First, many vertices redirect neighbors to the same central vertex, producing overly dense local structures that reduce graph diversity and navigability. Second, unnecessary reinsertion occurs when neighbors discarded by one vertex are concurrently reintroduced by others, essentially undoing pruning decisions. Third, exploration suppression emerges when uniform update behavior prevents the graph from discovering structurally diverse regions. Together, these effects degrade the constructed graph, undermining RNG pruning and ultimately reducing search accuracy. In addition, direct GPU implementations also suffer from irregular memory access and frequent dynamic allocations when maintaining per-vertex neighbor pools, further limiting scalability.

\subsection{Overall GRNND Framework}
To overcome the limitations of parallel RNN-Descent, we propose \textbf{GRNND}, a GPU-native redesign of the algorithm. GRNND preserves the iterative refinement principle of RNN-Descent but systematically restructures its update mechanism, memory model, and control flow to exploit massive parallelism.  
GRNND is composed of four key components: (1) disordered neighbor propagation to break premature convergence; (2) warp-level cooperative operations for fine-grained parallelism; (3) fixed-capacity double-buffered pools to avoid dynamic memory allocation; and (4) reverse edge sampling to maintain connectivity without excessive overhead. These components are integrated into a unified parallel pipeline.

As shown in Algorithm~\ref{alg:grnnd}, all vertices initialize in parallel with $S$ random neighbors (lines 3–5). The algorithm runs $T_1$ outer iterations, each consisting of $T_2$ warp-level update rounds. In every round, all vertices apply disordered propagation (lines 8–10), which samples candidate pairs at random and applies RNG-based redirection. After each outer iteration (except the last), reverse edges are inserted for a sampled subset (ratio $\rho$) to maintain connectivity and diversity (lines 13–15).
Throughout, warp-level primitives and fixed-size double buffers ensure efficient computation, safe concurrent updates, and regular memory access.

\subsection{Disordered Neighbor Propagation}
To avoid premature convergence caused by parallel ascending-order updates, we introduce a disordered neighbor propagation strategy tailored for GPU execution. Unlike the CPU implementation, which processes candidates in strictly ascending distance order, our approach evaluates them in randomized order. This breaks fixed update patterns and reduces symmetry across concurrent updates. For each vertex $v$, random neighbor pairs $(n_i, n_j)$ are sampled from its candidate pool and the RNG criterion is applied:  
\begin{equation}
\begin{aligned}
    &n_{\text{close}} = \arg\min_{x \in \{n_i, n_j\}} d(v, x), \ n_{\text{far}}   = \arg\max_{x \in \{n_i, n_j\}} d(v, x), \\
    &\text{if } d(n_i, n_j) < \max(d(v, n_i), d(v, n_j)) 
    \Rightarrow n_{\text{far}} \mapsto N_{n_{\text{close}}},
\end{aligned}
\label{eq2}
\end{equation}
where $v$, $n$, $d$, and $N$ denote the vertex, neighbor, distance, and candidate list, respectively.

\begin{algorithm}
\caption{UPDATE\_NEIGHBORS\_PARALLEL}
\label{alg:update_neighbors_parallel}
\begin{algorithmic}[1]
\STATE \textbf{Input:} $V$, $R$, $pool_1[v], pool_2[v]$ \textcolor{gray}{//double-buffered $N_v$}

\FORALL{$v \in V$ \textbf{in parallel}}
    \FOR{random neighbor pairs $(n_i, n_j) \in pool_1[v]$}
        \STATE $d_{ij} \leftarrow \text{WARP\_DISTANCE}(n_i, n_j)$
        \IF{$d_{ij} < \max(d(v, n_i), d(v, n_j))$}
            \STATE $n_{\text{close}} \leftarrow$ \text{closer}$(n_i, n_j)$, $n_{\text{far}} \leftarrow$ \text{farther}$(n_i, n_j)$
            \STATE \text{WARP\_INSERT}($n_{\text{far}}$, $n_{\text{close}}$, $pool_2[n_{\text{close}}]$)
            \STATE mark $n_{\text{far}} \leftarrow \bot$ 
        \ENDIF
    \ENDFOR

    \FORALL{$n \in pool_1[v]$}
        \IF{$n \neq \bot$}
            \STATE \text{WARP\_INSERT}($n$, $v$, $pool_2[v]$)
        \ENDIF
    \ENDFOR
\ENDFOR

\STATE \textbf{Clear} $pool_1[\ ]$ \textcolor{gray}{// reset $N_v$ for next round}
\STATE \textbf{Swap} $pool_1[\ ] \leftrightarrow pool_2[\ ]$ \textcolor{gray}{// swap pointers}
\end{algorithmic}
\end{algorithm}
\begin{figure*}[ht]
  \centering
  \includegraphics[width=0.9\linewidth]{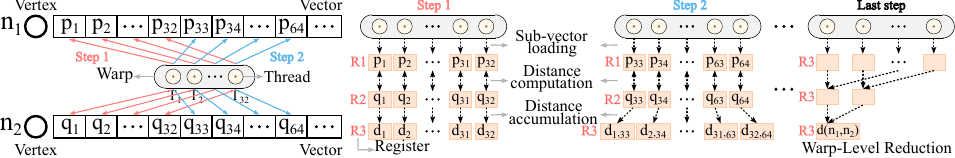}
  \caption{Warp-level distance computation: a single warp loads sub-vectors from two aligned inputs, computes squared differences in parallel, and performs warp-wide reduction.}
  \label{fig:warp_distance}
    \Description{A diagram illustrating warp-level distance computation. A single warp loads segments of two input vectors, element-wise computes their squared differences in parallel across threads, and then performs a warp-wide reduction to accumulate the total distance.}
\end{figure*}

This disordered strategy preserves the logic of the RNG criterion while introducing controlled randomness to break the synchronization that hinders convergence. As illustrated in Figure~\ref{fig:parallel_disordered}, the vertices can evaluate distant candidates (e.g., 4 and 5) before their closest neighbor (e.g., vertex 2). In such cases, redirection may occur toward previously unseen nodes, encouraging exploration of structurally diverse regions. Since each vertex executes randomized updates independently, global update trajectories naturally diverge. This mitigates synchronization, reduces unnecessary reinsertion, and improves pruning effectiveness. Over time, the graph converges more stably and exhibits greater structural diversity.

Algorithm~\ref{alg:update_neighbors_parallel} summarizes the parallel execution. In each round (line 3), the vertices simultaneously sample candidate pairs (line 4), calculate distances using warp-level primitives (line 5), and apply the RNG condition. If satisfied, the farther neighbor is redirected to the closer one (lines 6-10).

Although randomization may cause some close neighbors to be retained in later rounds rather than immediately,  repeated iterations and fixed double-buffered pools ensure their eventual preservation, yielding a stable balance between exploration and refinement and enabling reliable convergence with greater structural diversity. 

\subsection{Warp-Level Cooperative Operations} 
GRNND adopts a warp-level cooperative execution model to perform all stages of neighbor updates, including distance computation, deduplication, and insertion, as shown in lines 5, 8, and 14 of Algorithm~\ref{alg:update_neighbors_parallel}. Instead of assigning one thread per vertex or per neighbor, this often leads to control flow divergence and irregular memory access. GRNND delegates the full update process of each vertex to a single warp. This design is better aligned with the GPU hardware model and ensures efficient execution.
\begin{algorithm}
\caption{WARP\_DISTANCE$(n_i, n_j)$}
\label{alg:warp_distance}
\begin{algorithmic}[1]
\STATE \textbf{Input:} vectors $x_i$, $x_j$ of dimension $D$
\STATE \textbf{Output:} Euclidean distance $d(x_i, x_j)$

\STATE $tid \leftarrow \text{thread\_id\_in\_warp()}$, $sum \leftarrow 0$

\FOR{$d = tid; d < D; d += 32$}
\STATE $sum \leftarrow sum + (x_i[d] - x_j[d])^2$
\ENDFOR

\FOR{$offset = 16; offset \geq 1; offset = offset / 2$}
    \STATE $sum \leftarrow sum + {\_shfl\_down}(sum, offset)$
\ENDFOR
\RETURN $\sqrt{sum}$ \textcolor{gray}{// Only lane 0 holds final result}
\end{algorithmic}
\end{algorithm}
\begin{figure}[ht]  
  \centering
  \includegraphics[width=0.9\linewidth]{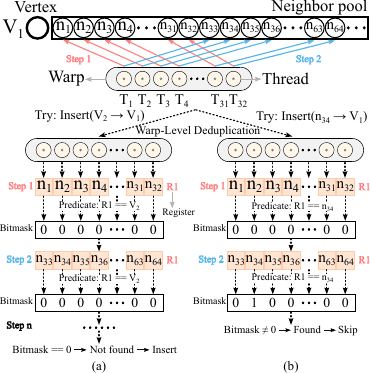}
  \caption{Efficient warp-level deduplication for candidate insertion using ballot function. (a) Insertion Allowed: $V_2$ not found in $V_1$’s candidate set; (b) Insertion Skipped: $n_{34}$ already exists in $V_1$’s candidate set.}
  \label{fig:ballot}
    \Description{Two subfigures illustrating warp-level deduplication using the CUDA ballot function. In subfigure (a), vertex V2 is not found in V1’s candidate set, so the insertion is allowed. In subfigure (b), candidate n34 is already present in V1’s candidate set, so the insertion is skipped. The diagram shows how threads collectively determine duplicates during insertion.}
\end{figure}

A critical performance stage is high-dimensional distance computation. As illustrated in Algorithm~\ref{alg:warp_distance}, GRNND assigns a single warp to compute the distance between two vectors. Each vector is divided into segments, and all warp threads collaborate to process one segment in parallel (lines 3-6). Within each segment, every thread loads a single dimension from both vectors, computes the squared difference, and accumulates the result locally (line 5). Once all segments are processed, a warp-wide reduction aggregates the partial results to produce the final distance (line 8). 
As illustrated in Figure~\ref{fig:warp_distance}, the computation has three stages: Step $1$, all warp threads collaboratively load and process the first segment in parallel; Steps $2$ to $n-1$, the warp sequentially iterates over the remaining segments in the same manner; Last step, a warp-wide reduction aggregates the partial sums into the final distance.

Warp-level cooperation also extends to neighbor insertion. To prevent redundant entries in the candidate list, GRNND uses warp-level ballot operations to perform set membership checks. As shown in Figure~\ref{fig:ballot}, each warp loads a segment of the pool into registers and compares the candidate neighbor against existing entries. The primitive ballot returns a bitmask indicating which threads detected a match. If no match is found, the warp continues scanning the next segment; otherwise, insertion is skipped. This strategy enables fast, register-level filtering without requiring global synchronization.
When a new neighbor is accepted, the warp either inserts it into an available slot or replaces an existing neighbor if the pool is full. In the replacement case, the warp uses max-reduction and ballot operations to locate the farthest neighbor, which is replaced only if the new candidate is closer. All insertions are performed using atomic operations to ensure correctness under concurrent access.
\begin{algorithm}
\caption{WARP\_INSERT($n_{\text{far}}, pool_{\text{close}}$)}
\label{alg:warp_insert}
\begin{algorithmic}[1]
\STATE $tid \leftarrow \text{thread\_id\_in\_warp()}$, $mask \leftarrow 0$

\STATE \textcolor{gray}{// Step 1: Warp-level deduplication}
\FOR{$i = tid; i < R; i += 32$}
\STATE $mask \leftarrow mask ~|~ \_ballot(pool_{\text{close}}[i] == n_{\text{far}})$
\ENDFOR
\IF{$mask \neq 0$}
    \RETURN \textcolor{gray}{// Already exists, skip insertion}
\ENDIF

\STATE \textcolor{gray}{// Step 2: Direct insertion if pool not full}
\STATE $count \leftarrow$ number of valid entries in $pool_{\text{close}}$
\IF{$count < R$}
    \STATE $pool_{\text{close}}[count] \leftarrow n_{\text{far}}$
    \RETURN
\ENDIF

\STATE \textcolor{gray}{// Step 3: Pool full, replace farthest neighbor if closer}
\STATE $max\_dist \leftarrow -\infty$, $rep\_idx \leftarrow -1$

\FOR{$i = tid; i < R; i += 32$}
    \STATE $d \leftarrow d(n_{\text{close}}, pool_{\text{close}}[i])$
    \IF{$d > max\_dist$}
        \STATE $max\_dist \leftarrow d$, $rep\_idx \leftarrow i$
    \ENDIF
\ENDFOR

\STATE $max\_dist \leftarrow \text{warp\_max}(max\_dist)$

\IF{$d(n_{\text{close}}, n_{\text{far}}) < max\_dist$ \AND $tid == rep\_idx$}
    \STATE $pool_{\text{close}}[tid] \leftarrow n_{\text{far}}$
\ENDIF

\end{algorithmic}
\end{algorithm}

As detailed in Algorithm~\ref{alg:warp_insert}, GRNND implements warp-level insertion through a structured three-stage process. Lines 2 to 8 perform warp-level deduplication using ballot primitives; if the candidate neighbor already exists, the insertion is skipped. Lines 10 to 14 handle direct insertion when the pool is not yet full. If the pool is full, lines 16 to 26 perform a warp-level max-reduction to locate the farthest neighbor and replace it only if the new candidate is closer.

By encapsulating neighbor updates within warp-local routines, GRNND achieves:
(1) uniform control flow with minimal divergence,
(2) fully coalesced and efficient memory access, and
(3) safe updates without inter-warp synchronization.
This design transforms the inherently sequential update logic of RNN-Descent into a scalable, deterministic, and parallel-safe routine.

\subsection{Fixed-Capacity Double-Buffered Pool}
\label{sec:buffering}
In CPU-based RNN-Descent, each vertex maintains a dynamic neighbor pool that is iteratively sorted, deduplicated, and truncated to keep the top-$R$ neighbors. This relies on sequential updates and safe, fine-grained memory access. On GPUs, however, concurrent vertex updates cause race conditions and synchronization overhead, while per-vertex dynamic arrays in global memory incur heavy costs from frequent allocations and atomic operations.
To address these issues, GRNND adopts a fixed-capacity double-buffered pool design. Each vertex is assigned two static buffers of size $R$: one for reading and one for writing. During iteration $t$, the read buffer holds the current neighbor set, while the write buffer accumulates filtered updates. 

Figure~\ref{fig:serial_parallel_compare} illustrates the update procedure: for each neighbor pair $(n_i, n_j)$ in the read buffer (pool 1), the RNG condition is applied. If the condition is not satisfied, the farther neighbor is redirected and inserted into the write buffer (pool 2) of the closer vertex. After all pairwise evaluations, the surviving neighbors from pool 1 are also copied into pool 2, subject to the same filtering rules. Once the update completes, the pointers of the two buffers are swapped: pool 2 becomes the new pool 1 for the next iteration. This behavior is captured in lines 12 to 19 of Algorithm~\ref{alg:update_neighbors_parallel}, where the retained neighbors are reinserted and the buffers are cleared and swapped. This double-buffered design eliminates dynamic memory allocation and fine-grained synchronization, while enabling deterministic, conflict-free, in-place neighbor updates under parallel execution. By preserving iteration-level consistency, it stabilizes convergence and becomes a key building block of GRNND’s GPU pipeline, supporting scalable and lock-free neighbor updates under massive parallelism.

\subsection{Reverse Edge Sampling}
Following disordered neighbor propagation, the graph already exhibits high structural diversity and exploratory connectivity. In this setting, inserting reverse edges for all neighbors incurs unnecessary overhead with diminishing returns. To balance graph quality and efficiency, GRNND adopts a reverse edge sampling strategy.
For each vertex $v$ with a neighbor set $N_v$ of size $k$, reverse edges are inserted only for the top $\rho \cdot k$ neighbors, where $\rho \in (0, 1)$ is a configurable sampling ratio. This approach takes advantage of the randomness introduced by disordered propagation, ensuring sufficient structural perturbation without requiring additional randomization during edge insertion.
Compared to exhaustive reverse insertion, this lightweight strategy reduces memory and computation overhead while using atomic writes only for the sampled subset to minimize contention and ensure correctness. At the same time, it effectively controls edge growth, preserves topological diversity, and supports convergence to high-quality graph structures at lower computational cost.

\section{EXPERIMENTS}
\label{sec_experiment}
\subsection{Experimental Setup}
\input{Exp_version2_4090_Ada/fig2}
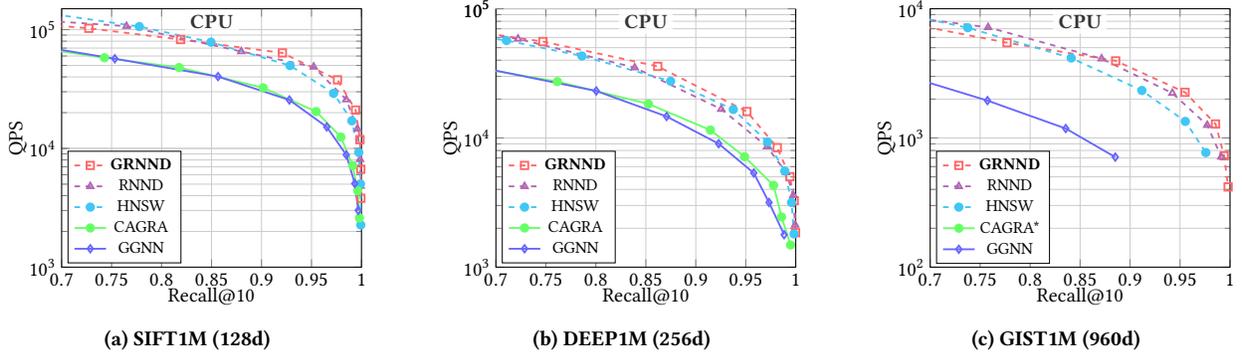
\begin{figure*}[t]
\begin{subfigure}{0.32\textwidth}
\centering
\begin{tikzpicture}[scale=0.7]
\begin{scope}
\begin{groupplot}[
  group style={
    group size=1 by 1,      
    vertical sep=25pt,
  },
  width=\textwidth, 
  scale only axis,
  grid=both,
  grid style={solid, gray!40},
  tick label style={font=\Large},
  title style={anchor=north,yshift=-6pt,font=\Large, fill=white,draw=none,fill opacity=0.8},
]
\nextgroupplot[
  ymode=log,
  xmin=0.70, xmax=1.00,
  xtick={0.70,0.75,0.80,0.85,0.90,0.95,1.00},
  ymin=1000, ymax=150000,
  ytick={1000,10000,100000},
  ylabel={QPS},
  xlabel={Recall@10},            
  legend style={at={(0.02,0.45)},anchor=north west},
  ylabel style={yshift=-5pt, font=\Large},
  xlabel style={yshift=5pt, font=\Large},
  title={\textbf{CPU}}, 
]
\addplot[smooth, tension=0, line width=1pt, mark=square,color=white!40!red, style=dashed, mark options={solid}] plot coordinates {
(0.6627, 113817.33)  
(0.7272, 102698.41)  
(0.819, 82571.95)  
(0.9209, 63701.42)  
(0.976, 37810.94)  
(0.9941, 21043.06)  
(0.9985, 11852.25)  
(0.9993, 6673.94)  
(0.9994, 3804.19)  
};
\addlegendentry{\textbf{GRNND}}  
\addplot[smooth, tension=0, line width=1pt, mark=triangle,color=white!40!violet, style=dashed, mark options={solid}] plot coordinates {
(0.6644, 121940.83)  
(0.7649, 106998.53)  
(0.8797, 65539.96)  
(0.9522, 48593.81)  
(0.9848, 25782.85)  
(0.9960, 14617.20)  
(0.9986, 8102.84)  
(0.9992, 4825.38)  
};
\addlegendentry{RNND}
\addplot[smooth, tension=0, line width=1pt, mark=*,color=white!40!cyan, style=dashed, mark options={solid}] plot coordinates {
(0.6048, 172225.10)
(0.7777, 106211.03)
(0.8494, 78507.66)
(0.9285, 49855.22)
(0.9721, 29092.19)
(0.9904, 17076.05)
(0.9973, 9251.40)
(0.9990, 5023.01)
(0.9993, 2267.49)
};
\addlegendentry{HNSW}  
\addplot[smooth, tension=0, line width=1pt, mark=*,color=white!40!green] plot coordinates {
(0.5555, 99358.42) 
(0.7428, 58033.07)
(0.8176, 47992.70)
(0.9016, 32402.70)
(0.9546, 20449.65)
(0.9794, 12458.99)
(0.9912, 7170.44)
(0.9962, 4396.30)
(0.9980, 2599.00)
};
\addlegendentry{CAGRA}
\addplot[smooth, tension=0, line width=1pt, mark=diamond,color=white!40!blue] plot coordinates {
(0.6709, 74054.22)
(0.7532, 56921.56)
(0.8565, 40285.55)
(0.9279, 25568.81)
(0.9653, 15163.80)
(0.9848, 8826.19)
(0.9933, 5061.53)
(0.9968, 3005.26)
};
\addlegendentry{GGNN}
\end{groupplot}
\end{scope}

\begin{scope}[shift={(0,-155pt)}] 

\end{scope}
\end{tikzpicture}
\caption{SIFT1M (128d)}
\end{subfigure}
\begin{subfigure}{0.32\textwidth}
\centering
\begin{tikzpicture}[scale=0.7]
\begin{scope}
\begin{groupplot}[
  group style={
    group size=1 by 1,      
    vertical sep=25pt,
  },
  width=\textwidth, 
  scale only axis,
  grid=both,
  grid style={solid, gray!40},
  tick label style={font=\Large},
 title style={anchor=north,yshift=-6pt,font=\Large, fill=white,draw=none,fill opacity=0.8},
]
\nextgroupplot[
  ymode=log,
  xmin=0.70, xmax=1.00,
  xtick={0.70,0.75,0.80,0.85,0.90,0.95,1.00},
  ymin=1000, ymax=100000,  
  ytick={1000,10000,100000},   
  extra y ticks={100000},       
  extra y tick labels={$10^{5}$},
  ylabel={QPS},
  xlabel={Recall@10},            
  legend style={at={(0.02,0.45)},anchor=north west, font=\normalsize},
  xlabel style={yshift=5pt, font=\Large},
  ylabel style={yshift=-5pt, font=\Large},
  title={\textbf{CPU}}, 
]
\addplot[smooth, tension=0, line width=1pt, mark=square,color=white!40!red, style=dashed, mark options={solid}] plot coordinates {
(0.648, 71963.95)  
(0.747, 55730.4)  
(0.8619, 35812.16)  
(0.9511, 16012.20)  
(0.9815, 8412.18)  
(0.9942, 5014.06)  
(0.9985, 3269.2)  
(0.9998, 1848.75)  
};
\addlegendentry{\textbf{GRNND}}  
\addplot[smooth, tension=0, line width=1pt, mark=triangle,color=white!40!violet, style=dashed, mark options={solid}] plot coordinates {
(0.6169, 76407.78)
(0.7221, 58448.33)  
(0.8389, 35042.21)  
(0.9255, 16718.14)  
(0.9713, 8561.54)  
(0.9904, 5396.06)  
(0.9976, 3606.35)  
(0.9994, 2123.69)  
};
\addlegendentry{RNND}
\addplot[smooth, tension=0, line width=1pt, mark=*,color=white!40!cyan, style=dashed, mark options={solid}] plot coordinates {
(0.6680, 63402.51)
(0.7106, 56774.14)
(0.7861, 43108.22)
(0.8748, 27511.00)
(0.9375, 16629.32)
(0.9718, 9309.85)
(0.9891, 5540.40)
(0.9958, 3167.59)
(0.9983, 1810.01)
};
\addlegendentry{HNSW}  
\addplot[smooth, tension=0, line width=1pt, mark=*,color=white!40!green] plot coordinates {
(0.6891, 34059.92)
(0.7615, 27313.34)
(0.8526, 18344.52)
(0.9146, 11508.45)
(0.9488, 7155.45)
(0.9778, 4297.05)
(0.9858, 2447.48)
(0.9948, 1484.94)
};
\addlegendentry{CAGRA}
\addplot[smooth, tension=0, line width=1pt, mark=diamond,color=white!40!blue] plot coordinates {
(0.6881, 34665.82)
(0.8004, 23110.50)
(0.8708, 14663.55)
(0.9229, 9025.36)
(0.9580, 5357.70)
(0.9733, 3165.75)
(0.9884, 1782.97)
};
\addlegendentry{GGNN}
\end{groupplot}
\end{scope}

\begin{scope}[shift={(0,-155pt)}] 

\end{scope}
\end{tikzpicture}
\caption{DEEP1M (256d)}
\end{subfigure}
\begin{subfigure}{0.32\textwidth}
\centering
\begin{tikzpicture}[scale=0.7]
\begin{scope}
\begin{groupplot}[
  group style={
    group size=1 by 2,      
    vertical sep=25pt,
  },
  width=\textwidth, 
  scale only axis,
  grid=both,
  grid style={solid, gray!40},
  tick label style={font=\Large},
 title style={anchor=north,yshift=-6pt,font=\Large, fill=white,draw=none,fill opacity=0.8},
]
\nextgroupplot[
  ymode=log,
  xmin=0.70, xmax=1.00,
  xtick={0.70,0.75,0.80,0.85,0.90,0.95,1.00},
    ymin=100, ymax=10000,  
    ytick={100,1000,10000},  
  ylabel={QPS},
  xlabel={Recall@10},            
  legend style={at={(0.02,0.45)},anchor=north west, font=\normalsize},
  xlabel style={yshift=5pt, font=\Large},
  ylabel style={yshift=-5pt, font=\Large},
  title={\textbf{CPU}}, 
]
\addplot[smooth, tension=0, line width=1pt, mark=square,color=white!40!red, style=dashed, mark options={solid}] plot coordinates {
(0.6302, 9045.16)  
(0.7769, 5464.92)  
(0.8856, 3950.61)  
(0.955, 2256.5)  
(0.9855, 1282.78)  
(0.9941, 727.29)  
(0.9981, 417.85)  
};
\addlegendentry{\textbf{GRNND}}  
\addplot[smooth, tension=0, line width=1pt, mark=triangle,color=white!40!violet, style=dashed, mark options={solid}] plot coordinates {
(0.6233, 9759.46)  
(0.7580, 7182.34)  
(0.8715, 4108.86)  
(0.9423, 2223.36)  
(0.9772, 1256.89)  
(0.9912, 714.39)  
};
\addlegendentry{RNND}
\addplot[smooth, tension=0, line width=1pt, mark=*,color=white!40!cyan, style=dashed, mark options={solid}] plot coordinates {
(0.6212, 11061.59)
(0.7376, 7133.77)
(0.8411, 4157.40)
(0.9118, 2330.72)
(0.9553, 1343.29)
(0.9758, 770.66)
};
\addlegendentry{HNSW}  
\addplot[smooth, tension=0, line width=1pt, mark=*,color=white!40!green] plot coordinates {
(0.6, 100) 
};
\addlegendentry{CAGRA*}
\addplot[smooth, tension=0, line width=1pt, mark=diamond,color=white!40!blue] plot coordinates {
(0.6719, 3096.21) 
(0.7573, 1946.82)
(0.8356, 1183.25)
(0.8852, 710.91)
};
\addlegendentry{GGNN}
\end{groupplot}
\end{scope}

\begin{scope}[shift={(0,-155pt)}] 

\end{scope}
\end{tikzpicture}
\caption{GIST1M (960d)}
\end{subfigure}
\caption{Evaluating QPS vs. recall: Fixed construction settings with method-specific search tuning to assess the performance of the constructed indices.  
\footnotesize\textit{$^*$On GIST1M, CAGRA is unavailable due to high memory demand.}}
\Description{QPS–recall evaluation against state-of-the-art methods: construction configuration and search algorithm are fixed, while search parameter are tuned per method to assess the search performance of the constructed indices. \footnotesize\textit{On GIST1M, CAGRA is unavailable due to high memory demand.}}
\label{fig_3}
\end{figure*}
\input{Exp_version2_4090_Ada/sort}
\input{Exp_version2_4090_Ada/sample}
\input{Exp_version2_4090_Ada/T1T2}
To comprehensively evaluate the performance of GRNND, we compare it against several state-of-the-art GPU-based ANNS algorithms as well as strong CPU baselines. Our primary focus is on index construction time, measured under comparable or equivalent search quality.
We include the following representative algorithms for comparison.
\begin{itemize}
\item CAGRA\cite{ootomo2024cagra}: A refinement-based GPU method that follows a build-then-prune paradigm.
\item GANNS\cite{yu2022gpu}: A GPU-optimized direct construction algorithm with enhancements inspired by NSW.
\item GGNN\cite{groh2022ggnn}: A hierarchical GPU-based method designed for efficient graph generation.
\item RNN\cite{ono2023relative}: The fastest known CPU implementation of RNN-Descent with high construction quality.
\item HNSW\cite{malkov2018efficient}: A widely used CPU-based multi-layer graph structure optimized for search accuracy.
\item GRNND: Our proposed GPU-parallel algorithm of RNN-Descent.
\end{itemize}

To ensure fair comparisons, we use a unified evaluation metric for index construction time. For flat graphs (non hierarchical), we first construct and save the index, then use the same search algorithm to evaluate Recall@10 and construction time under equivalent parameters. For hierarchical methods, we use their native search routines and match the neighbor pool sizes to maintain consistency. In addition to the core comparisons, we conducted ablation studies to isolate the impact of individual GRNND components and analyze how different hyperparameters affect both index construction time and search performance.
All experiments were carried out in two hardware configurations: an NVIDIA RTX 4090 with an Intel Core i9-14900KF CPU and an NVIDIA RTX 6000 Ada with an AMD Ryzen Threadripper PRO 7965WX CPU. The inclusion of both a consumer GPU and a professional GPU ensures the robustness of the results across different hardware environments.

\subsection{Comparison with Other Methods}
Figure~\ref{fig_2} compares index construction time at matched recall across all methods on RTX 4090 and RTX 6000 Ada, respectively. For fairness, we fix the search algorithm and search parameter. The parameter corresponds to the maximum size of the candidate list maintained during graph traversal, while each method tunes only its construction configuration to reach a comparable Recall@10.
GRNND consistently outperforms state-of-the-art GPU-based ANNS algorithms on both platforms, achieving comparable or better search quality with significantly lower construction time.
On RTX 4090, GRNND achieves $2.6$ to $3.9\times$ speedup over GANNS, $3.1$ to $5.0\times$ over CAGRA, $21.5$ to $48.3\times$ over GGNN, $27.1$ to $38.7\times$ over CPU-based RNN and $31.9$ to $49.8\times$ over HNSW. On RTX 6000 Ada, GRNND achieves $2.4$ to $4.1\times$ speedup over GANNS, $4.3$ to $5.7\times$ over CAGRA, $30.6$ to $51.7\times$ over GGNN, $17.8$ to $25.1\times$ over CPU-based RNN and $22.7$ to $26.2\times$ over HNSW. The results demonstrate consistent advantages on both consumer- and professional-grade GPUs, thereby validating robustness across heterogeneous hardware environments.

Notably, GRNND demonstrates strong robustness to increasing dimensionality. Its construction time scales proportionally with vector dimension, maintaining high throughput even on high-dimensional datasets. This is primarily enabled by GRNND’s warp-level cooperative computation, which parallelizes both distance calculations and neighbor updates with minimal divergence and memory contention.
In the GIST1M dataset (960 dimensions), GRNND achieves its largest performance margin, highlighting its scalability. In contrast, GGNN and CAGRA show degraded performance under high dimensionality. GGNN suffers from excessive construction time and lower search accuracy due to ineffective hierarchical graph construction. CAGRA frequently fails to complete due to memory exhaustion before pruning and refinement phases.
GRNND’s superior performance is attributed to two key design elements: its fixed-capacity double-buffered pool, which ensures consistent and conflict-free memory access, and its disordered neighbor propagation strategy, which enhances structural diversity and avoids the convergence failures caused by globally synchronized updates. Together, these techniques allow GRNND to build high-quality graphs efficiently, even under challenging high-dimensional conditions.

Figure~\ref{fig_3} shows CPU query performance using a unified search algorithm. The comparison is conducted across graphs built by each method under comparable construction configurations. 
%
%
After construction, all graphs are evaluated with the same search algorithm, and QPS–recall curves are obtained by varying the search parameter that controls the candidate list size.
GRNND achieves search quality comparable to that of CPU-based RNN-Descent, confirming that our GPU-parallel design preserves graph quality. 
Moreover, the graphs generated by GRNND achieve search performance slightly higher than HNSW and substantially better than GPU-based GGNN and CAGRA. Since GRNND achieves the shortest construction time among all methods, this highlights the superiority of our GPU-parallel algorithm. Here, GANNS is excluded from the comparison since its index cannot be evaluated with a search algorithm independent of its native query implementation.
%

\subsection{Ablation Study}
Figures~\ref{sort} and~\ref{sample} evaluate the impact of various optimization strategies in GRNND on both RTX 4090 and RTX 6000 Ada. Both the fixed-capacity double-buffered structure and warp-level cooperative updates are always enabled as part of the baseline design.

Figure~\ref{sort} compares the disordered execution strategy with ascending and descending sorting orders on the SIFT1M, DEEP1M, and GIST1M datasets. We include descending sorting to provide a comprehensive understanding of how the update order influences structural diversity and search accuracy. 
The observed trends are consistent across both GPUs.
Notably, the RTX 6000 Ada exhibits slightly longer construction time than the RTX 4090 under identical configurations. This is primarily due to differences in clock frequency and power configuration, as the professional-grade Ada architecture is tuned for sustained stability and large memory capacity rather than maximizing raw throughput, in contrast to the consumer-oriented RTX 4090.

Ascending sorting, which prioritizes nearest neighbors, tends to lead to synchronous convergence traps during parallel execution. When all vertices update their neighbors simultaneously, early convergence restricts topological diversity, which is particularly problematic in parallel GPU environments.
On the other hand, descending sorting emphasizes distant neighbors, promoting better global exploration. However, this increases computational overhead, especially on high-dimensional datasets like GIST1M, as it requires excessive exploratory updates to maintain diversity.
In contrast, disordered execution strikes a more balanced approach. By randomly selecting neighbor pairs for evaluation based on the RNG condition, disordered execution maintains structural diversity without incurring the overhead seen in descending sorting. While slightly slower than ascending sorting, disordered execution is much faster than descending sorting and achieves superior search accuracy.
Thus, disordered execution avoids the convergence trap inherent in ascending sorting and mitigates the overhead of descending sorting. This makes it an effective strategy for high-dimensional, large-scale graph construction.

Figure~\ref{sample} illustrates the effect of varying the back-edge sampling ratio on the performance of GRNND across datasets. The overall trend of the RTX 4090 and RTX 6000 Ada is consistent, highlighting the robustness of the design. A low sampling ratio produces faster index construction, but reduces connectivity and structural diversity. This degradation is particularly pronounced on high-dimensional datasets such as GIST1M, where sparse neighborhoods amplify the loss of reverse edges. Increasing the sampling ratio improves graph quality and search accuracy by incorporating more reverse connections, yet the gain diminishes once sufficient connectivity has been established. Beyond this point, further increases in the ratio no longer yield meaningful accuracy improvements but monotonically prolong construction time. This overhead arises because additional reverse neighbors enlarge the candidate pool and incur higher refinement costs during each iteration.
Overall, the results demonstrate that a moderate sampling ratio (around $0.6$) achieves the best balance between efficiency and quality. It preserves sufficient structural diversity to sustain high recall while avoiding the excessive refinement overhead associated with higher ratios. Thus, careful tuning of the sampling ratio is essential for constructing high-quality graphs efficiently, especially in high-dimensional settings.

\subsection{Parameter Sensitivity}
Figure~\ref{T1T2} examines the influence of the outer iteration count ($T_1$) and the inner iteration count ($T_2$) on graph construction across the three benchmark datasets. The observed tendencies are consistent for both RTX 4090 and RTX 6000 Ada.
The effect of $T_2$ exhibits a strong dependence on the dimensionality of the data. Increasing $T_2$ generally improves graph refinement and increases search accuracy, but its impact on construction time diverges between datasets. In SIFT1M, where the points are densely distributed, neighborhood structures converge quickly in a few refinement steps. Additional $T_2$ iterations only introduce redundant comparisons and insertions, leading to a nearly monotonic increase in construction time. In contrast, on higher-dimensional datasets such as DEEP1M and GIST1M, where the neighborhoods are sparser and the initial quality of the candidate is lower, moderate increases in $T_2$ accelerate the formation of high-quality candidate pools in early iterations. This reduces redundant neighbor updates in later outer loops, and in some cases even shortens the overall construction time.

The role of $T_1$ is comparatively uniform: more outer iterations inevitably prolong the construction time, since each iteration entails a full graph refinement. However, the benefits are dimension-dependent. For low-dimensional datasets (SIFT1M, DEEP1M), one or two outer iterations already yield a near-optimal graph, with further rounds providing only marginal accuracy gains at significant time cost. In contrast, on the high-dimensional GIST1M, increasing $T_1$ continues to improve graph quality and recall, as repeated refinement is necessary to escape local optima and overcome the extreme sparsity of high-dimensional space.

In summary, both GPUs exhibit the same performance–quality trade-off: shallow optimization cycles suffice for low-dimensional data, while high-dimensional datasets demand deeper iterations to achieve competitive accuracy. The joint tuning of $T_1$ and $T_2$ is therefore essential to balance efficiency and quality, and consistent cross-platform trends further validate the generality of the proposed design.

\section{RELATED WORK}
\label{sec_releatedworks}
Early GPU efforts mainly focused on the search stage, such as SONG~\cite{zhao2020song} and FAISS~\cite{johnson2019billion}. As datasets scaled, index construction emerged as the primary bottleneck, motivating fully GPU-based solutions.

GNND\cite{wang2021fast}, by Wang et al., is a GPU adaptation of NN-Descent. It uses selective neighbor updates to reduce global memory traffic and optimized sampling and distance computation to achieve $100$ to $250\times$ speedup over single-threaded NN-Descent, outperforming GPU baselines like GGNN and FAISS by $2.5$ to $5\times$.

CAGRA\cite{ootomo2024cagra}, by Ono et al., is a GPU-Native framework for both construction and querying. It constructs an initial K-NN graph and refines it through pruning. With warp-segmented execution, forgettable hash tables, and topology-aware pruning, it significantly reduces memory and maintenance costs—achieving 2–27× speedup over HNSW in graph building.

GANNS\cite{yu2022gpu}, by Yu et al., is a GPU-parallel graph construction and search framework optimized for NSW and NSG. It uses CUDA-based concurrent node expansion and local atomics with global consistency checks to manage connections. Experiments demonstrate that it outperforms SONG and earlier GPU-HNSW methods in both speed and accuracy.

GGNN\cite{groh2022ggnn}, by Groh et al., partitions the datasets into subsets, constructs subgraphs, and merges them hierarchically. It uses thread-block granularity for task division and employs hierarchical scheduling and unified neighbor pools to reduce contention. It surpasses FAISS in build speed.


\section{CONCLUSION}
\label{sec_conclusion}
This paper presents GRNND, the first GPU-parallel algorithm of Relative NN-Descent for efficient approximate nearest neighbor (ANN) graph construction. GRNND preserves the core refinement principles of RNN-Descent, but systematically redesigns its execution flow, memory organization, and update strategy to match the massively parallel architecture of modern GPUs. The main contributions include a disordered neighbor propagation strategy that mitigates synchronous convergence traps, a warp-cooperative update mechanism that ensures both efficiency and correctness under concurrency, and a fixed-capacity double-buffered pool that avoids the overhead of dynamic allocation. These innovations jointly enable GRNND to deliver high-quality graphs with substantially reduced construction time. Extensive experiments on multiple datasets and hardware platforms demonstrate that GRNND consistently outperforms both CPU- and GPU-based baselines, validating its robustness across dimensionalities and system configurations.
Future work will focus on extending GRNND to broader application scenarios, such as billion-scale indexing, multi-GPU or distributed deployments, and integration into end-to-end retrieval and analytics pipelines. Another promising direction is further system-level optimization, including memory management and scheduling policies, to maximize throughput on emerging GPU architectures.

\begin{acks}
This work was supported by the China Scholarship Council. Additional funding was provided by Cross-ministerial Strategic Innovation Promotion Program (SIP) on “Integrated Health Care System” Grant Number JPJ012425, JST, and CREST Grant Number JPMJCR22M2, Japan.
\end{acks}


\bibliographystyle{ACM-Reference-Format}
\bibliography{main}

\end{document}